\documentclass[preprint]{aastex}

\newcommand{\phunit}{photons cm$^{-2}$ sr$^{-1}$ s$^{-1}$}
\newcommand{\ergunit}{ergs cm$^{-2}$ sr$^{-1}$ s$^{-1}$}
\newcommand{\lya}{Ly~$\alpha$}
\newcommand{\lyb}{Ly~$\beta$}
\newcommand{\fdegree}{\,\hbox{$.\!\!^\circ$}}

\shorttitle{Upper limits on \ion{O}{6} Emission From {\it Voyager}
Observations}
\shortauthors{Murthy et al.}

\begin{document}

\title{Upper limits on \ion{O}{6} Emission From {\it Voyager} Observations}
\author{Jayant Murthy}
\affil{Indian Institute of Astrophysics, Koramangala, Bangalore 560 034}
\email{jmurthy@yahoo.com}

\author{R. C. Henry and R. L. Shelton}
\affil{The Johns Hopkins University, Baltimore, MD 21218}

\and

  \author{J. B. Holberg}
\affil{Lunar and Planetary Laboratory, University of Arizona, Tucson, AZ 85721}

\begin{abstract}

We have examined 426 {\it Voyager} fields distributed across the sky  for
\ion{O}{6} ($\lambda \lambda$~1032/1038~\AA ) emission from the Galactic
diffuse interstellar medium. No such emission was detected
in any of our observed fields. Our
most constraining limit was a 90\%
confidence upper limit of 2600 \phunit\ on the doublet emission in
the direction
(l, b) = (117.3, 50.6). Combining this with an absorption line
measurement in nearly the same direction allows us to place
  an upper limit of 0.01 cm$^{-3}$  on the electron density of the hot
gas in this
direction. We have placed 90\% confidence upper limits of less than
or equal to 10,000 \phunit\
on the \ion{O}{6} emission in 16 of our 426 observations.

\end{abstract}


\section{Introduction}

There have been many detections of ultraviolet resonance line absorption
by highly ionized, presumably hot, gas in the Galactic halo
(e.g., Sembach \& Savage 1992, Hurwitz \& Bowyer 1996), but
only three claimed detections of ultraviolet
resonance line emission from this gas.
First, Martin \&  Bowyer (1990)
reported detections of the (unresolved) \ion{C}{4} ($\lambda \lambda$
1548/1550 \AA )
emission in 4 out of their 8
lines of sight with a maximum strength of $(7.3 \pm 0.9) \times
10^{-8}$ \ergunit.
They also detectected O III] ($\lambda$ 1663 \AA ) emission at about half the
intensity and much lower significance. Second, out of the ten Hopkins
Ultraviolet Telescope (HUT) targets which could be profitably
used for studies of the Galactic halo, \citet{dixon96}
detected \ion{O}{6} emission in four directions at
levels on the order of  $4 \times 10^{-7}$ \ergunit\ ($2.1 \times 10^4$
\phunit), with 2 $\sigma$ upper limits of roughly the same level
on the other six targets. Finally, there have been three recent observations 
of \ion{O}{6} emission using the {\it FUSE} satellite all at a level of about only
5000 photons cm$^{-2}$ sr$^{-1}$ s$^{-1}$: \citet{shelton00} at (l, b) = 
(315.00, -41.33); and \citet{dixon01} at (l, b) = (284.2, 74.5)
and (l, b) = (57.6, 88.0).

The most constraining upper limit is the {\em MINISAT-01 }  90\%
confidence upper limit of
$2.5 \times 10^{-8}$ \ergunit\ (1200 \phunit) from
\citet{edelstein99} with earlier limits of about $1.6 \times 10^{-7}$
\ergunit\ (7600 \phunit) from \citet{korpela98} and
\citet{holberg86}. Note that
all values cited for the \ion{O}{6} emission are for the integrated
emission over both lines of the doublet.

\citet{murthy99} have reprocessed 17 years (1977 -- 1994)
of data from the {\em Voyager} 1 and~2 archives with a focus on the
continuum emission due to dust scattering. In the present paper, we will
discuss limits, from the same data set, on \ion{O}{6} (1032/1038~\AA )
line emission from the ISM. Although new instruments are now
providing important results, the {\em Voyager} data are still the only
source of information on the \ion{O}{6} emission over many different
lines of sight.

\section{Observations and Data Analysis}

The two {\em Voyager} spacecraft were launched in 1977 and have
taken FUV (500  - 1700 \AA) spectra of astronomical objects ever
since. Each spacecraft includes a Wadsworth-mounted objective
grating spectrometer (UVS) with a field of view of 0\fdegree1 x 0\fdegree87
and a spectral resolution of 38 \AA\  for aperture filling diffuse
sources. A full description of the UVS instruments and the {\em
Voyager} mission
is given by \citet{holberg92}.

The data processing is described in \citet{murthy99}.
Because we were only interested in the diffuse background,
all other observations (planets, stars and nebulae) were
discarded.  The remaining data consist of 426 observations
of diffuse background.
The \ion{O}{6} doublet ($\lambda \lambda$ 1032/1038~\AA)
is clearly visible in the {\em Voyager} spectra of bright sources such
as supernovae remnants \citep{blair95} and the Eridanus
superbubble \citep{murthy93}, where the doublet is much
brighter than the heliospheric hydrogen \lyb\
($\lambda $ 1026~\AA) emission.  However, the
\ion{O}{6} emission from the diffuse
halo gas is much less than the \lyb\ emission on whose wings it lies.
Fortunately,
because the Lyman lines are optically thick, the \lyb/\lya\ ratio
is constant throughout the heliosphere and we can use the \lya\
line to scale the \lyb\ line. We determined the ratio between the
two lines using UVS observations in which only the
heliospheric lines were present and then
used this empirical ratio to scale the \lyb\ line in each observation
(see \citet{murthy99}
for a full description of this procedure). We subtracted this scaled
\lyb\ intensity
from the observed spectrum and determined the \ion{O}{6} upper limit
from the remainder.

Because the \lyb\ line is at almost the same position as the \ion{O}{6} line,
there is a tradeoff between their respective derived intensities.
Note, however, that the difference in the central wavelengths of the \lyb\ and \ion{O}{6} emission is
large enough (1 resolution element) that \ion{O}{6} cannot fully, or even largely, replace the
\lyb\ contribution. In our procedure, we have restricted the heliospheric \lyb/\lya\ ratio
to fall between empirically
determined limits; if, on the other hand, we allow the \lyb/\lya\ ratio
to vary freely, our limits on the \ion{O}{6} emission will be correspondingly poorer.
We have carried out this exercise for each of our
targets but, because of the varying amount of heliospheric emission,
cannot compare points on an individual basis. In
general, our limits go up (become less constraining) by a factor
of  2 -- 3. Thus our best upper limit rises from 2600 \phunit\ to
8500 \phunit\ and the number of pointings with \ion{O}{6}
limits under 50,000 \phunit\ drops to 127 from 244.
(Details of the individual targets are available from the authors
on request.)

\citet{edelstein00} have claimed that \citet{holberg86} and \citet{murthy99},
have significantly underestimated the errors in the {\em Voyager}
data. As the data analysis
in this work rests heavily on the earlier papers, particularly that
of \citet{murthy99},
we are compelled to address these criticisms. A careful reading of
the Edelstein et al. paper
shows that there are virtually no differences between their results
and ours, despite their
claims. From their Table 2 Edelstein et al. obtain a 1~$\sigma$
uncertainty of 125 \phunit\ \AA$^{-1}$, identical
to the limit claimed by \citet{holberg86}.
However, they also obtain a residual of 320 \phunit\ \AA$^{-1}$
whereas \citet{holberg86} found
a null signal. This difference can be traced to Edelstein et al. estimating
two large numbers from a figure in \citet{holberg86}, subtracting the
two and claiming that as
the residual. If they had actually used the original data in digital
format, the correct procedure,
their results would have agreed exactly with \citet{holberg86}.
Importantly for this work, and contrary to their claims, \citet{edelstein00}
have shown that the counting errors in our analysis procedure are reasonable.

Because the signal in the {\em Voyager} spectra is dominated by the
RTG particle background
(due to radioactivity in the radioisotope thermoelectric generator)
and the scattered \lya, we have explored the possibility
that systematic
errors in the subtraction of the two components are affecting our
\ion{O}{6} limits.
Should there be a feature in either of these components
coincidentally at the position of \ion{O}{6}, we
would expect the \ion{O}{6} limits to be correlated with that
component, because the
strength of that feature would be necessarily correlated with the
level of the continuum (the sum
of the RTG and \lya\ contributions). Over the 17 years of {\it
Voyager} observations, both the RTG level
and the \lya\ emission declined: the former because of the decline in
the radioactivity of the plutonium
power source and the latter because of the increasing distance of the
spacecraft from the Sun.
Thus, if there were any significant systematic errors associated with
the background subtraction, our
\ion{O}{6} limits would be strongly correlated with the year of observation.
No such effect is detectable in our data (shown in the case of the
\lya\ emission in Figure 1),
implying that systematic effects
due to the subtraction of the RTG and \lya\ backgrounds are
unimportant.

We can demonstrate empirically that our quoted error bars are
reasonable through
a listing of each of the errors in one of our targets (Table 1). Note
that we have arbitrarily
chosen the location with our most constraining \ion{O}{6} limit. We
have listed in Table 1 the integrated counts under the \ion{O}{6} line in the
total spectrum and in each of the modeled components of the raw {\em Voyager}
data: the RTG spectrum; the \lya\ template; \lyb\ emission; and the
diffuse continuum (due to dust-scattered starlight). The poisson
errors are also listed, with the RTG and total
errors reflecting the fact that each RTG event generates 3 counts
\citep{holberg86}. From these errors we then calculate a total
uncertainty, assuming
uncorrelated errors. For this target we obtain a 1 $\sigma$ uncertainty
of  308 counts corresponding to a signal of 1400 \phunit .
This is entirely consistent with the 90\%
confidence level of 2600 \phunit\  that we derived
using our modeling procedure (the modeling is described in detail in \citet{murthy99}). Our quoted
uncertainties have taken into
account all the statistical errors and come from a $\chi^2$ minimization
according to the procedure described by \citet{lampton76}. Essentially, we
changed the value of the \ion{O}{6} emission, while allowing the
other parameters to
vary freely through the allowed parameter space, until the $\chi
^{2}$ emission rose to
unacceptable levels.

\section{Results and Discussion}

We detect no \ion{O}{6} emission in any of 426 UVS observations of
the diffuse radiation
field but do set upper limits on such emission in each direction.
The best of these limits is 2600 \phunit\
($5.0 \times 10^{-8}$ \ergunit ) in the \ion{O}{6} resonance line doublet
in the direction (l, b) =
(117.3, 50.6). This direction is quite close to HD
121800 (l, b = 113.0,
49.8 , spectral type B1.5~V, distance = 2.2~kpc) towards which
\citet{hurwitz96} obtained a \ion{O}{6}
column density of $1.1 \times 10^{14}$ cm$^{-2}$ using ORFEUS. Using
these values and
Equation 5 of \citet{shull94}, and confining the
temperature range to that for which the fraction of oxygen atoms
in the \ion{O}{6} state is within 10\% of its maximum value in
collisional ionization equilibrium plasma (T = $2.2 - 6.4 \times 10^5$
K ---  \citet{shapiro77}), we find an upper limit on the electron
density of less than
0.010 cm$^{-3}$.  Assuming that the emitting
gas has a solar abundance of helium atoms and that the hydrogen
and helium are fully ionized, there will be 1.9 particles per
electron and thus the thermal pressure will be less than 12,000 K
cm$^{-3}$, close to the thermal pressure of 15,000 K cm$^{-3}$ in the Local
Bubble derived by \citet{snowden98} from observations of the
1/4 keV soft X-ray flux seen by ROSAT.

The 94 locations in which we set 90\% confidence upper limits of
better than $5 \times 10^{-7}$ \ergunit\ (25,000 \phunit) are plotted
in Figure 2
and those in which we set limits of better than $2 \times 10^{-7}$
\ergunit\ ($\approx$ 10,000 \phunit)
are listed in Table 2. Several of our observations are near the locations
observed by \citet{dixon96} using HUT and we both set similar
upper limits in those (with our {\em Voyager} limits in general being
more constraining).
Only in their Target 3 (UGC 5675; l = 218.2, b = 56.4) do
we obtain inconsistent results,
with \citet{dixon96} quoting a flux of 23,000 $\pm$ 6000  \phunit\
while we place a 90\% upper
limit of 10$^4$ \phunit at (l, b) = (216.8, 55.3) --- about 1$^\circ$
away. Of course, it is entirely
possible that there are truly spatial variations of this scale in the ISM.

We also have several observations near the four high latitude
locations where Martin \&  Bowyer (1990) detected \ion{C}{4} emission
but in none can we do more than say that the \ion{O}{6}/C~IV ratio
is not inconsistent with the theoretical ratios reported from a
variety of physical conditions (e.g. cooling flows: \citet{edgar86};
shock heated gas: \citet{hartigan87}; fountains: \citet{shapiro93};
halo supernova
remnants: \citet{shelton98}).

\section{Conclusion}

Very recent results concerning galactic diffuse  \ion{O}{6} emission
include the {\em FUSE} detections
by \citet{shelton00} and \citet{dixon01} at a level of 5000 \phunit\
and the {\em MINISAT-01} all-sky upper limit of 1200 \phunit\
by \citet{edelstein99}. Combined with the present {\em Voyager}
upper limits, it appears that much of the sky has an \ion{O}{6}
emission of significantly less than 10,000 photons cm$^{-2}$
sr$^{-1}$ s$^{-1}$.
Only the 4 HUT detections of \citet{dixon96} show higher fluxes. A
mission dedicated
to the observation and mapping of faint line emission from the
Galactic halo would surely yield bountiful results.

\acknowledgments

Much of this work was done while JM was at the Johns Hopkins
University and was supported by
NASA grants NAG5-2398 and NAG5-2299. An anonymous referee contributed
greatly to a discussion of the
error analysis in this work. We thank Doyle Hall and Matthew Earl for their
contributions to the data analysis.

\clearpage

\begin{figure}
\plotone{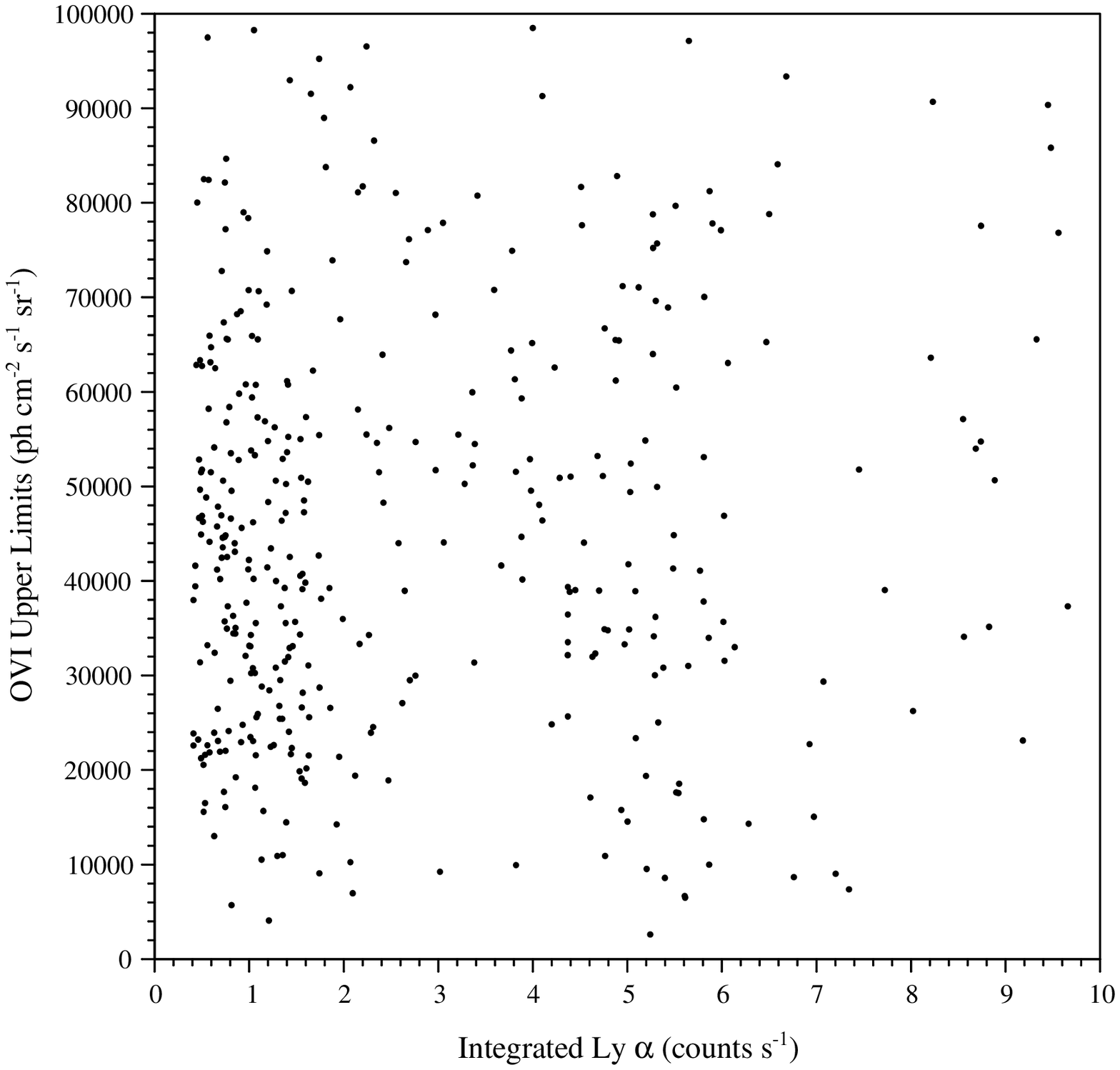}
\caption{We have plotted the derived \ion{O}{6} upper limits against
the integrated counts in the \lya\ line. There is no correlation
between the two.  Such a correlation might be expected if systematic
errors were
an important contributor to the \ion{O}{6} limits. Similar plots are
obtained when the \ion{O}{6} limits are plotted against the relative RTG
strength or the diffuse continuum - none of the
different components are correlated, indicating that systematic errors are
unimportant in our analysis. \label{fig1}}
\end{figure}

\clearpage

\begin{figure}
\plotone{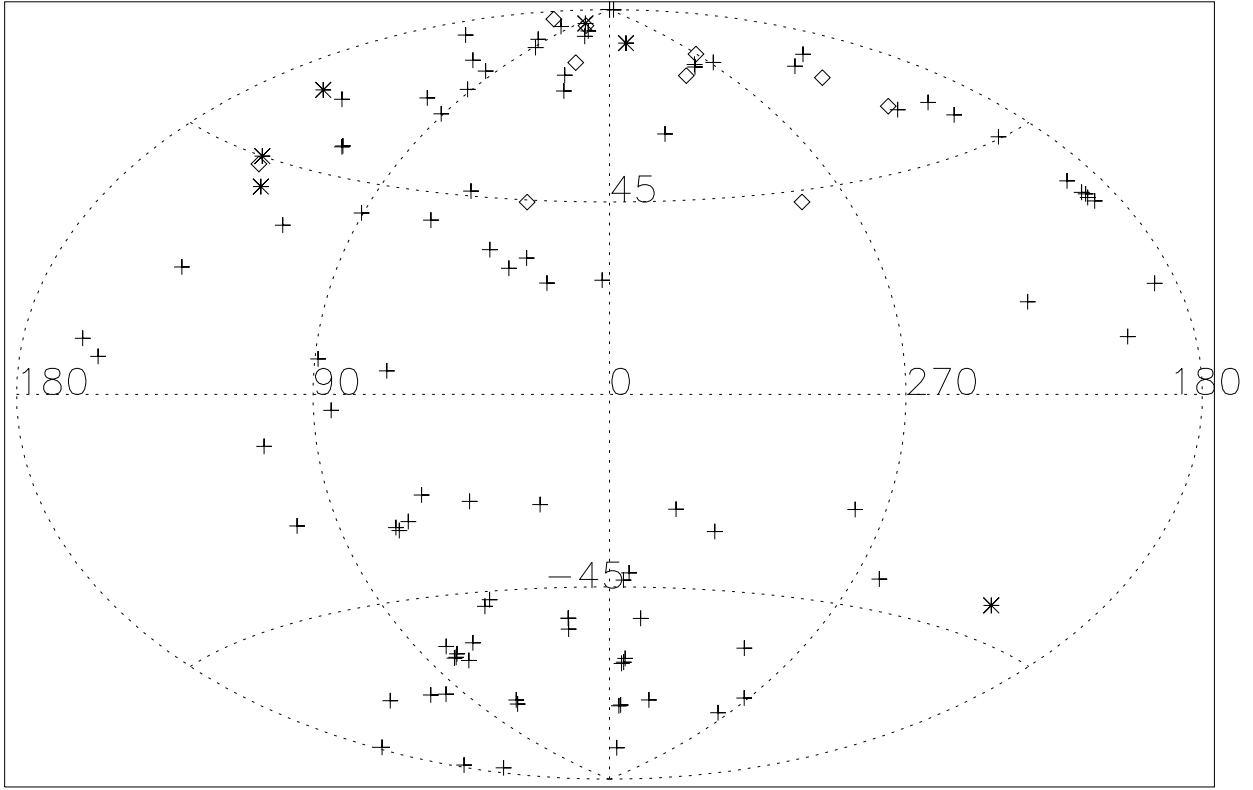}
\caption{All of the {\em Voyager} observations in which we were able
to set upper limits of less than $5 \times 10^{-7}$ \ergunit\ (25,000
\phunit) are
plotted as plus signs on an Aitoff map of the sky with the
origin at the center and 180$^\circ$ at the left. The Dixon et al.
(1996) targets are plotted as diamonds and the C~IV detections
of Martin \&  Bowyer (1990) are plotted as asterisks. In one
direction in common (see text), we place a 90\% confidence limit that
is about half the
claimed detection by Dixon et al.; however, given both sets of
uncertainties and the different locations, we cannot rule out their
claimed value. \label{fig2}}
\end{figure}

\clearpage
\begin{deluxetable}{rcc}
\tablecaption{Statistical Limits on the Modeled Components \label{tbl-1}}
\tablewidth{0pt}
\tablehead{
\colhead{ } & \colhead{Total Counts}   & \colhead{Poisson Error}    }
\startdata
Total & 113000 & 194\\
RTG & 91200 & 174\\
\lya & 2300 & 48\\
\lyb & 11700 & 108\\
Diffuse Continuum & 11800 & 109\\
Total Error & & 308\\ \enddata
\end{deluxetable}

\clearpage
\begin{deluxetable}{ccc}
\tablecaption{Best {\em Voyager} \ion{O}{6} Upper Limits \label{tbl-2}}
\tablewidth{0pt}
\tablehead{
\colhead{l} & \colhead{b}   & \colhead{90\% Confidence Upper Limit on
\ion{O}{6} Emission}  \\
\colhead{degrees} & \colhead{degrees}   & \colhead{\phunit}   }
\startdata
117.3   &             50.6   &              2,600\\
272.5    &            -67.4  &              4,100\\
67.8      &           5.2    &              5,700\\
60.3       &          -22.5  &              6,500\\
117.3       &         50.8   &              6,700\\
200.7        &        9.6    &              7,000\\
71.6          &       -59.6  &              7,400\\
189.6          &      32.3   &              8,600\\
91.1            &     61.4   &              8,700\\
115.7            &    72.6   &              9,000\\
32                &   70.5   &              9,100\\
331.7              &  60.5   &              9,200\\
99.3                & 80.3   &              9,500\\
225.7  &              68.3   &              9,900\\
190     &             33.3   &              10,000\\
216.8    &            55.3   &              10,000\\
346.6     &           -52.3  &             11,000\\
  \enddata
\end{deluxetable}

\end{document}